\input amstex
\magnification=1200
\documentstyle{amsppt}
\NoRunningHeads
\pagewidth{142 mm}
\vcorrection{-15 mm}
\topmatter

\title The addition of the lower level to spectrums of matrix and scalar 
components of d=2 SUSY Hamiltonian \endtitle
\author S B Leble, A V Yurov \endauthor
\address {\rm Technical University 
of Gdansk, Faculty of Technical Physics and Applied Mathematics
ul. G.Narutowicza 11/12 80-952 Gdan'sk-Wrzeszcz Poland
236041, Theoretical Physics Department, Kaliningrad State
University, Al.Nevsky St., 14, Kaliningrad, Russia;}
\endaddress
\email
leble\@mifgate.pg.gda.pl 
\endemail
\abstract
Supersymmetrical quantum--mechanical system is consider in the case of
d=2. The problem of  addition of the lower level to spectrums of matrix and
scalar components of d=2 SUSY Hamiltonian is investigated. It is shown that
in the case, the level E=0 may be degenerate. The multi--dimensional scalar
Hamiltonians with energy spectra coinciding up to finite number of discrete
levels are constructed.

\endabstract
\endtopmatter
\document

{\bf 1.Introduction}

Supersymmetric quantum mechanics realizes the quantum description of systems
with double degeneracy of energy levels. When d=1, the supersymmetry has
absorbed in itself the one--dimensional factorization method being 
intrinsically
connected with the Darboux transformation (DT) [1], [2]. The DT group together
two Hamiltonians $h_0$ and $h_1$ with equivalent spectra:
$$h_0=q^+q+E_0,\qquad h_1=qq^++E_0,\qquad q\equiv\frac {d}{dx} - (ln\varphi)',
\eqno(1)$$
where $q^+$ is the Hermitian conjugate with $q$ operator, $\varphi$ is
solution of equation $h_0\varphi=E_0\varphi$ ($support$ $function$). It easy 
to see that
$h_0$ and $h_1$ are intertwined by $q$ and $q^+$
$$q\,h_0=h_1\,q,\qquad h_0\,q^+=q^+\,h_1,\eqno(2)$$
and therefore
$$\psi^{(1)}=q\,\psi,\qquad \psi=q^+\,\psi^{(1)},\eqno(3)$$
if
$$h_0\psi=E_1\psi,\qquad h_1\psi^{(1)}=E_1\psi^{(1)},\eqno(4)$$
and  and $\varphi^{(1)}=\varphi^{-1}$.
For the brevity, the normalizing multipliers in (3) are omitted.

DT  give us the way to
construct one--dimensional potentials with arbitrary preassigned discrete
spectrums. For example if support function $\varphi(E_0;x)$ is wave
function (w.f.) of the ground state $h_0$, then the discrete spectrum of
$h_1$ is coinciding with the spectrum of $h_0$ without lower level $E_0$ [3]. 
It is possible to add the level $E_0$ (missing in spectrum of $h_0$) to 
spectrum of $h_1$. To this end it is sufficient to exploit such $\varphi$ that:
$$\varphi\to+\infty,\qquad x \to \pm \infty,$$
and $\varphi$ is positive definite function for all values $x$. 
I is convenient to choose 
$\varphi$ as:

$$\varphi=\lambda \varphi_++(1-\lambda)\varphi_-,\eqno (5)$$
where $\varphi_+$ and $\varphi_-$ are positive definite functions with the
following asymptotic behavior:
$$\varphi_{\pm}\to\cases + \infty,&\text {for $x\to\pm\infty$}\\
 0,&\text {for $x\to\mp\infty$},\endcases\eqno(6)$$
and $\lambda$ is real parameter lying in interval [0,1]. If
$0<\lambda <1$, then the level $E_0$ is the lower of spectrum $h_1$. If
$\lambda=0$ or $\lambda=1$, the level $E_0$ is missing in spectrums of $h_0$
and $h_1$ and spectrums of these Hamiltonians are coinciding with
each other.

All this relates to the one-dimensional supersymmetric quantum mechanics
based on the following commutation relations:

$$[Q,H]=[Q^+,H]=0,\qquad \{Q,Q^+\}=H,\eqno(7)$$
where
$$Q=q\sigma_+,\qquad Q^+=q^+\sigma_-,\qquad H=diag(h_0-E_0,h_1-E_0),\eqno(8)$$
and $\sigma_{\pm}=(\sigma_1\pm i\sigma_2)/2$, $\sigma_{1,2}$ are Pauli matrices

Let's remind that $q$ and  $q^+$ are -- bosonic and $\sigma_-$,$\sigma_+$ -- 
fermionic
operators of creation-annihilation. If the spectra of $h_0$ and $h_1$ differ
by the  only level,
then (7) corresponds to the exact supersymmetry. If the level $E_0$ is absent
at spectra of both operators - the 
supersymmetry is said broken. It is easy to see that the level
$E_0$ cannot {\bf simultaneously} be presented at the spectrum of $h_0$ 
and at spectrum of
$h_1$. 
It means that if the lowest level at the spectum of the supersymmetric 
hamiltonian is zero, it is degenerate

At this work we consider the case of two-dimension supersymmetric quantum
mechanics. The explicit form of operators that satisfy th algebraic relations 
(7) at d = 2 is determined by the expressions: 
$$
Q=\pmatrix
0&0&0&0\\
q_1&0&0&0\\
q_2&0&0&0\\
0&q_2&-q_1&0
\endpmatrix, \qquad
Q^+=\pmatrix
0&q^+_1&q^+_2&0\\
0&0&0&q^+_2\\
0&0&0&-q^+_1\\
0&0&0&0
\endpmatrix,\eqno(8)
$$
$$
H=diag(h_0-E_0,\tilde h_{lm}-2\delta_{lm}E_0,h_1-E_0),\eqno(9)
$$
where
$$h_0=q_m^+q_m+E_0,\qquad h_1=q_mq_m^++E_0, \qquad \tilde h_{lm}\equiv h_{lm}
+H_{lm}-E_0\delta_{lm},\eqno(10)$$
by the way 
$$h_{lm}=q_lq_m^++E_0\delta_{lm},\qquad H_{lm}=p_lp_m^+
+E_0\delta_{lm},\eqno(11)$$
$q_l=\partial_l-\partial_l(\ln\varphi)$, $p_l=
\varepsilon_{lk}q_k^+$, $\varepsilon_{lk}$ --
is the antisimmetric tensor
 $\partial_l\equiv \partial/\partial x^l$, with indices
l=1, 2 and the sum by repeated indices is implied.

In contrast to  d=1, there no any association between spectra $h_0$ and $h_1$ 
in general case.
For the illustration of the last statement let us consider
The Coulomb potential (the example is taken from the article 
[3]) $u=-\alpha\,r^{-1}$, $\alpha>0$. It is easy to verify that
$u^{(1)}=+\alpha\,r^{-1}$.

At the example , ad hoc, the hamiltonian  $h_1$ has no discrete spectrum at all
whereas
at  d=1 the spectrum of $h_1$ should be obtained by the deleting
of the lowest level from the spectrum of $h_0$.
The assertion should clear  that when $d>1$ we
have no formulas expressing wave functions of 
$h_1$via WF of $h_0$, that could be similar to one-dimensional case.
The existence of such formulas means the existence 
of connections between spectra
 $h_0$ and $h_1$. However it is not forbidden the existence of potentials
of a special form that allow the connections between spectra. 
Moreover there could be expressions that connect wave functions of
the corresponding hamiltonians
$h_0,1$ that do not relay to a physical spectrum.
As we shall see both possibilities allow a realization.

The general coupling between the spectra exist for pairs
$h_0$, $h_{lm}$ and $h_1$, $H_{lm}$.
Really, taking into account that
 $h_1$ may be represented as  $h_1=p_m^+p_m+E_0$, it is easy to verify
the validity of the intertwine relations
: 
$$q_lh_0=h_{lm}q_m ,\qquad p_lh_1=H_{lm}p_m ,$$

$$h_0q_l^+=q_m^+h_{ml} ,\qquad h_1p_l^+=p_m^+H_{ml},\eqno(12)$$
That means the presence of such coupling. By the same formulas the
operator $\tilde h_{lm}$. 
is inter(laced)twined with
$h_0$ and $h_1$
Its spectrum coincides with the spectrums of
the scalar hamiltonians 
excluding may be the level $E_0$.

At the works [3, 4, 5] it was studied the supersymmetry defined by operators
(8)--(9), with assumption that $\varphi$  is a wave function 
of the basic state of the hamiltonian
$h_0$. As it was shown at the cited papers
such a choice of
$\varphi$ leads to the assertion that the level $E_0$ is absent at 
the physical parts of spectra 
of $\tilde h_{lm}$ and $h_1$, or to unbroken supersymmetry. In this paper we 
study the inverse problem -- the addition of the level $E_0$, that is absent
at spectrum of $h_0$to the spectrums of both other operators.
In this part the general approach should be developed 
that give a possibility to realize
this procedure. At the Sec. 3 we consider
the explicit example of potentials 
(scalar and matrix) with the additional level.
It will be shown that the resulting supesymmetric hamiltonian
possess a twice degenerated level with
$E=0$ --that we did not met for d=1 case in general
and for  d=2 within the  "level deleting" case. At the last section
we describe briefly the algorithm of d=2 widen supersymmetry model construction.

{\bf 2. The level addition}

Let $u=u(x,y)$  be a n integrable potential i.e. it is supposed
that we able to solve the Schr{\"o}dinger equation 
 $h_0\psi=E\psi$ explicitly for any spectral parameter value
 $E$, $h_0=-\Delta+u$. Unlike the one-dimensional case the potential
$$u^{(1)}=u-2\Delta\ln\varphi,\eqno(13)$$
(where $\varphi$ is $support$ $function$, is not integrable. Suppose that
the spectral parameter value $E_0$ lies below the basic state energy
of the hamiltonian $h_0$. The following question is important for us:
how the 
$support$ $function$ $\varphi$ should be chosen that the level $E_0$
appear at the physical part of the spectrums
of
$h_1$ and $\tilde h_{lm}$?

For a scalar hamiltonian the reply for this question
is not difficult.
Really
it is easy to verify that the function $\varphi^{-1}$ satisfy th e equation
$$h_1\frac {1}{\varphi}=E_0\frac {1}{\varphi},\eqno(14)$$
therefore it is enough to choose  $\varphi$ as positive function for all
x and y that grows exponentially in all directions in plane. 
The situation coincide literally with one-dimensional case
(if one don't consider the exited levels).

For the matrix hamiltonian it is necessary the separate consideration
First of all we note 
that if the function $\psi$ is the second solution of Schr{\"o} dinger equation
with $E_0$, than tha function 
$$\tilde \psi_m=q_m\psi,\eqno(15)$$
satisfy the equation
$$\tilde h_{lm}\tilde \psi=E_0\tilde \psi.\eqno(16)$$
Show now that for rapidly decreasing $\tilde \psi$ the representation (15)
is not only sufficient but necessary as well.

For the beginning we prove that the level $E_0$
belong to the spectrum of $\tilde h_{lm}$, iff the respecting
{\bf is normable} wave function $\tilde \psi_m$ satisfy the condition:
$$h_{lm}\tilde \psi_m=H_{lm}\tilde \psi_m=E_0\tilde \psi_m.\eqno(17)$$
Really let exists the function $\tilde \psi_m$ such that
$$\tilde h_{lm}\tilde \psi_m=
E_0\tilde \psi_m,\qquad (\tilde \psi_m,\tilde \psi_m)=1.\eqno(18)$$
Define the functions $\rho_m$ and $\sigma_m$ by equalities
$$\rho_m\equiv h_{lm}\tilde \psi_m,\qquad \sigma_m\equiv H_{lm}\tilde 
\psi_m.\eqno(19)$$
From (10)--(11) it follows that $\sigma+\rho=2E_0\tilde \psi$ (indices
omitted), i.e.
$$(\rho+\sigma,\rho+\sigma)=4E_0^2\eqno(20)$$
if $\rho$ and $\sigma$ - are normable. Otherwise one may 
check that
$$h_{mk}H_{kl}=H_{mk}h_{kl}=E_0\tilde h_{ml}\eqno(21).$$
From (21) it follows
$$h_{lm}\sigma_m=H_{lm}\rho_m=E_0^2\tilde \psi_l.\eqno(22)$$
Hence
$$(\tilde \psi_m,h_{lm}\sigma_l)=(h_{lm}\tilde \psi_m,\sigma_l)=
(\rho_m,\sigma_m)=E_0^2.\eqno(23)$$
Combining with (20) we obtain $(\rho-\sigma,\rho-\sigma)=0$, therefore
$\sigma_m=\rho_m=E_0\tilde \psi_m$. Finally from (19) we go to the equation
(17) $\blacksquare$.

Thus for the level $E_0$ lie at physical spectrum $\tilde h_{lm}$,
it is necessary to find a normable solution 
of (17). Let $\tilde \psi_m$ is such the function
. Acting for it by $\tilde h_{lm}$, we get the equation
$$q_m^+\tilde \psi_m=p_m^+\tilde \psi_m=0.\eqno(24)$$
It means that there exist two functions $\psi$ and $\psi^{(1)}$, such that
$$\tilde \psi_m=q_m\psi=p_m\psi^{(1)}\eqno(25)$$
and that do satisfy the equations
$$h_0\psi=E_0\psi,\qquad h_1\psi^{(1)}=E_0\psi^{(1)}.\eqno(26)$$
Solving  (25) with respect to $\psi^{(1)}$, we get the importany relation
that couple $\psi$ and $\psi^{(1)}$:
$$\psi^{(1)}=\frac {1}{\varphi} \int dx_k\varepsilon_{km}(\varphi\partial_m\psi-
\psi\partial_m\varphi),\eqno(27)$$
that is known as Moutard transformation
[6]. It remains to note that from the established connections between
 $\tilde \psi_m$, $\rho_m$ and $\sigma_m$, the formula  (15) obviously follows.

Thus for the presence of the level $E_0$ in the spectrum of  $\tilde h_{lm}$
it should be pointed out two normable solutions
$\psi$, $\varphi$ of the Schr{\"o}dinger equation with apotential $u$ 
and the spectral parameter $E=E_0$, for the function $\tilde \psi_m=q_m\psi$ 
be normable.
At the next section we would illustrate this procedure by an example.

{\bf 3. Potentials with cylindrical symmetry.}

Let $\psi$ and $\varphi>0$ are the solutions described at the end of previous 
section
. For the construction of matrix potentials 
with the level $E_0$ it is convenient to introduce an auxiliary function
$$f\equiv \frac {\psi}{\varphi},\eqno(28)$$
that satisfies the equation
$$\partial_m(\varphi^2\partial_mf)=0.\eqno(29)$$
Then
$$\tilde \psi_m=\varphi\partial_mf.\eqno(30)$$

Consider the case when the seed potential 
possess the cylindrical symmetry 
$u=u(r)$. Integrating (29) and substituting in (30) one get
$$\tilde \psi_m=\frac {x_m}{r^2\varphi}.\eqno(31)$$
The normalizing integral of (31) converge
if $\varphi$ grows at infimity
as a power function with arbitrary index and at the vicinity of zero
it behave as $r^{-k}$, $k>0$. 
If one require that the asymptotic behavior
of $\varphi$
is determined by the conditions:
$$\varphi\to \cases r^a,&\text {for $x^2+y^2\to\infty$}\\
 r^b,&\text {for $x^2+y^2\to 0$},\endcases\eqno(32)$$
where $a>1$, $b<1$, then the normalizing integral 
of $\varphi^{-1}$ should converge as well.
This means that the level $E_0$ will present  at  the both spectrums
of the operators  $h_1$ and  $\tilde h_{lm}$ simultaneously.

In the reference [3] it was shown that the similar situation
cannot take place for the hamiltonians
$h_0$ and $\tilde h_{lm}$. It is easy to see the difference between these couples
. For example using as the $support$ $function$ $1/\varphi$,
it is possible to construct a new supersymmetric hamiltonian
$$\hat H=diag(h_1-E_0,\hat h_{lm}-2\delta_{lm}E_0,h_0-E_0).\eqno(33)$$
The operator $\hat h_{lm}$ differs from $\tilde h_{lm}$ by that it is 
intertwined with
$h_1$ not by the operators $p_m$, but the dual ones $q_m^+$. 
Respectively in its
spectrum the level $E_0$ do not exist
as for the rest of these operators spectrums coincide.
Note that such "equivalent by spectrum" matrix operators were considered
in the work [7].

The spectrum of the supersymmetrical Hamiltonian
(9), consist of the levels
$$\{E_i-E_0,E^{(1)}_i-E_0\},$$
where $E_i$, $E^{(1)}_i$ -- are the levels of the discrete spectrums
parts of 
 $h_0$ and $h_1$ -- respectively.
Now it is seen that if the condition (32) is satisfied then at the spectrum of
(9) there is twice degenerated level
 $E=0$, to which the following eigen functions correspond
$$
\Psi_1=\pmatrix
0\\
0\\
0\\
\frac {1}{\varphi}
\endpmatrix,\qquad
\Psi_2=\pmatrix
0\\
\varphi\partial_1f\\
\varphi\partial_2f\\
0
\endpmatrix.\eqno(34)
$$
Use the explicit form of the odd supersymmetric
operators (8), prove the validity of the known relations for wave functions
for the zero level
$$Q\Psi_{1,2}=Q^+\Psi_{1,2}=0.$$

As an example choose $\varphi=\exp(br)/r^k$, where $b, k>0$.
 Such function satisfy the necessary asymptotic
(32). As a result we obtain two scalar potentials 
of the hamiltonians
$h_0$ and  $h_1$:
$$u=\frac {k^2}{r^2}-\frac {b(2k-1)}{r},\qquad u^{(1)}=
\frac {k^2}{r^2}-\frac {b(2k+1)}{r}.\eqno(35)$$
The additional level corresponds to the energy 
$E_0=-b^2$. It may be verified by the note that the potentials are integrated
by means of degenerated Hypergeometric functions.
The discrete spectrums are determined by the formulas:
$$E_{_N}=-\frac {b^2(2k\mp 1)^2}{(1+2[N+\sqrt{m^2+k^2}])^2},\eqno(36)$$
where the sign "-" corresponds to $u$, and "+" -- to $u^{(1)}$, $N$ -- is 
the principal and $m$ -- magnetic
quantum numbers.

The constructed potentials are interesting as an example that exhibits 
a difference
between DT  in multidimensions and their unidimensional counterpart.
Specifically the comparison of spectrums of hamiltonians 
$h_0$ and
$h_1$ shows that the addition of the lowest level shifts all spectrum.
If consider the potentials (35), it could be seen that when
$$k=\frac {(N+1)^2-m^2}{2(N+1)},$$
the addition of the level $E_0=-b^2$, do not move the exited level
with the number $N$ and fixed $m$. I general the levels of the hamiltonian
$h_1$ go down in respect to levels of $h_0$. 
This displacement is maximal in the lowest part of the well and decrease as 
 $1/N^2$ in the higher part of the spectrum. In tutn the spectrum of the 
supersymmetric Hamiltonian (9) is double degenerated, including the level
$E=0$. Its normable  vacuum wave functions are given by the expressions
(34), and
$$\frac {1}{\varphi}=r^k\exp(-br),\qquad \partial_mf=
\frac {x_m}{(r\varphi)^2}.\eqno(37)$$

{\bf 4.Extended supersymmetry}

In the previous section we had demonstrated how to built up a
two-dimensional Hamiltonian with double degenerated level
$E=0$. One may obtain models with all degenerated levels
. Such thing is realized in models with extended symmetry
[8], [9]:
$$\{Q_i,Q_k\}=\delta_{ik}H,\qquad [Q_i,H]=0,\qquad i,k=1,...,N.\eqno(38)$$

Two hamiltonians to be defined: $H_1$ -- is determined by (9), and $H_2$,
differs from it by the permutation of  $h_0$ and $h_1$. We need 
the three operators more: $Q_1$ -- defined by (8) and
$$
Q_2=\pmatrix
0&0&0&0\\
-q^+_2&0&0&0\\
q^+_1&0&0&0\\
0&-q^+_1&-q^+_2&0
\endpmatrix, \qquad
B=\pmatrix
0&q_2&-q_1&0\\
-q_1&0&0&q^+_2\\
-q_2&0&0&-q^+_1\\
0&-q^+_1&-q^+_2&0
\endpmatrix, \eqno(39)
$$
that satisfy the commutation relations
$$Q^+_2B+BQ^+_1=Q_2B+BQ_1=BH_1-H_2B=0,\eqno(40)$$
$$H_1=\{Q^+_1,Q_1\}=B^+B, \qquad H_2=\{ Q^+_2,Q_2\}=BB^+.\eqno(41)$$
One can verify that the operators:
$$H(2)=diag(H_1,H_2), \qquad Q_1(2)=diag(Q_1,Q_2),\eqno(42)$$
$$
Q_2(2)=\pmatrix
0&0\\
B&0
\endpmatrix,\eqno(43)
$$
form the algebra (38) when $N=2$.

The introduced operators are 
the construction blocks for the building of extended 
supersymmetry matrices for any
$N$. It is possible to demonstrate that at every step 
there is an operator factorizing a superhamiltonian as in 
(41).	At the step $N$, the superhamiltonian
$$
H(N)=\pmatrix
B^+_{_N}B_{_N}&0\\
0&B_{_N}B^+_{_N}
\endpmatrix,\eqno(44)
$$
is factorized by the operators $B_{_{N+1}}=diag(B_{_N},B^+_{_N})$. 
In turn
$H(N+1)$, $B_k(N+1)$, ($k\le N$) are defined by the substitution
 $B_{_N}\to B_{_{N+1}}$ and
the addition of the new operator $Q_{_{N+1}}(N+1)$ with the new matrix structure
. At every step the dimension of matrices duplicates therefore the corresponded
algebra is realized by matrices $2^{N+1}\times 2^{N+1}$.
For example at $N=4$ there would be four operators $Q_i$ of the dimension 
$32 \times 32$ 
and the similar superhamiltonian
$H$:
$$H\equiv H(4)=diag(H_1,H_2,H_2,H_1,H_2,H_1,H_1,H_2),\eqno(45)$$
$$Q_1(4)=diag(Q_1,Q_2,Q_2,Q_1,Q_2,Q_1,Q_1,Q_2),\eqno(46)$$
$$
Q_s(4)=\pmatrix
0&0&0&0&0&0&0&0\\
B&0&0&0&0&0&0&0\\
B&0&0&0&0&0&0&0\\
0&B^+&-B^+&0&0&0&0&0\\
B&0&0&0&0&0&0&0\\
0&B^+&0&0&-B^+&0&0&0\\
0&0&B^+&0&-B^+&0&0&0\\
0&0&0&B&0&-B&B&0
\endpmatrix,\eqno(47)
$$
and also $s=2,3,4$. The nonzero elements of the three
corresponding operators are 
over the main diagonal
of the matrix (47).If as a basic scalar model one take 
the potential considered in the previous section it is obvious that
all levels including zero one are degenerate with the multiplicity
$2^N$.

{\bf REFERENCES}

1.Infeld., Hull T.E., Rev. Mod. Phys., 1951, {\bf 23}. p.21.

2.Darboux G. C.R.Acad.Sci Parus, 1882, {\bf 92}, p.1456.

3.Crum M.M. Quart. J. Math. Oxford. 1955 {\bf 6}, n 2, p.121.

4. Andrianov A. Borisov N. Ioffe M. Phys.Lett.A 1984 105, 19-22.

5. Adler Teor, Mat. Phys. 1994, {\bf 101} p.323, Russian.

6.V P Berezovoy and A I Pashnev  1987  Theor. and Math. Phys. 
{\bf 70}, 146,  Russian.

7.A A Andrianov, M V Borisov, M V Ioffe and M I Eides 1985 
 Phys.Lett.{\bf 109A}, p 143.

8.A A Andrianov,  M V Ioffe, Nishnianidze D.N. Theor. and Math. Phys. 
 {\bf 104},p.463, Russian. 

9.Moutard. Th.F. C.R.Acad.Sci.Paris. 1875. V.80 p.729.

10. Matveev V.B., Salle M.A. Darboux transformations and solitons. 
Springer-Verlag, Berlin, 1991.

11.Andrianov A.A.,Ioffe M. Phys.Lett. 1990,{\bf B255}, p.136.

12.V P Berezovoy and A I Pashnev  1988, Theor. and Math. Phys. 
 {\bf78}, p.136, Russian.

 13.Salle M.A., Yurov A.V. The factorization of two-dimensional 
 hamiltonians and algebra of extended supersymmetry. Preprint NMCAO
 N9306-3. S.- Petersburg,1993.

\enddocument
\bye